\documentclass[11pt,aps,amsmath,amssymb,nofootinbib,notitlepage,longbibliography]{revtex4-2}
\usepackage{amsmath,amssymb}
\usepackage{slashed}
\usepackage{graphicx}
\usepackage{bm}
\usepackage[top=1.0in,bottom=1.0in,left=1.0in,right=1.0in]{geometry}
\usepackage[colorlinks,linkcolor=blue,citecolor=blue]{hyperref}

\newcommand{\be}{\begin{equation}}
\newcommand{\ee}{\end{equation}}
\newcommand{\bea}{\begin{eqnarray}}
\newcommand{\eea}{\end{eqnarray}}
\newcommand{\ba}{\begin{eqnarray}}
\newcommand{\ea}{\end{eqnarray}}

\begin{document}

\title{Rapidity evolution  of the entanglement entropy in quarkonium:\\
parton and string duality}

\author{Yizhuang Liu}
\email{yizhuang.liu@uj.edu.pl}
\affiliation{Institute of Theoretical Physics, Jagiellonian University, 30-348 Kraków, Poland}

\author{Maciej A. Nowak}
\email{maciej.a.nowak@uj.edu.pl}
\affiliation{Institute
of Theoretical Physics and Mark Kac Center for Complex Systems Research,
Jagiellonian University, 30-348 Kraków, Poland}

\author{Ismail Zahed}
\email{ismail.zahed@stonybrook.edu}
\affiliation{Center for Nuclear Theory, Department of Physics and Astronomy, Stony Brook University, Stony Brook, New York 11794--3800, USA}



\begin{abstract}
We investigate the quantum entanglement in rapidity space of the soft gluon wave function of a quarkonium, in theories with non-trivial rapidity evolutions. We found that the rapidity evolution drastically changes the behavior of the entanglement entropy, at any given order in perturbation theory. At large $N_c$, the reduced density matrices that ``resum'' the leading rapidity-logs can be explicitly constructed,  and shown to satisfy Balitsky-Kovchegov (BK)-like evolution equations. We study their entanglement entropy in a simplified $1+1$ toy model,  and in
3D QCD. The entanglement entropy in these cases, after re-summation, is shown to saturate the Kolmogorov-Sinai bound of 1. Remarkably, in 3D QCD the essential growth rate of the entanglement entropy is found to vanish at
large rapidities, a result of kinematical ``quenching'' in transverse space. The one-body reduction of the entangled density matrix obeys a BFKL evolution equation,
which can be recast as an evolution in an emergent AdS space, at large impact-parameter and large rapidity. This observation allows the extension of the
perturbative wee parton evolution at low-x,  to a dual non-perturbative evolution of string bits in curved AdS$_5$ space,
with manifest entanglement entropy in the confining regime.
\end{abstract}

\maketitle

\section{Introduction}
Quantum entanglement
permeates most of our  quantum description of physical laws.
It follows from the fact that quantum states
are mostly superposition states, and two non-causally related measurements can be correlated,
as captured by the famed EPR paradox. A quantitative
measure of this correlation is given by the quantum entanglement entropy.
The entanglement entropy of quantum many body system and quantum field theory has been extensively explored in the literature~\cite{Srednicki:1993im,Calabrese:2004eu,Casini:2005rm,Hastings:2007iok,Calabrese:2009qy}.

In hadron physics, quantum entanglement is inherent to any hadron state, which is made more
spectacular on the light-front with luminal wave-functions. Unlike a generic quantum many body system, quantum field theory is
intrinsically multi-scale in nature which leads to non-trivial evolutions with respect to energy and rapidity scales. In particular, in 4D gauge theories,
under large boosts, the wave-functions pile more and more {\it small-$x$ partons} or {\it wee partons}~\cite{Feynman:1969wa} and results in nontrivial asymptotic behaviors in physical cross sections.

A comprehensive understanding of the small-$x$ asymptotics in 4D gauge theory,  is still not available in so far even in perturbation theory, but there are progresses. In weak coupling QCD, the rapidity evolution is  extensively discussed in the  literature,  leading to various evolutions equations such as the BFKL equation~\cite{Kuraev:1977fs,Balitsky:1978ic,Lipatov:1996ts} or the BK equations~\cite{Balitsky:1995ub,Kovchegov:1999yj}. In strong coupling QCD,  they are identified with {\it string bits}~\cite{Susskind:1993ki,Susskind:1993aa,Thorn:1994sw},
and well described by a stringy evolution equation in the double limit of strong gauge coupling and large $N_c$. The general idea is, the small-$x$ partons behave very differently with respect to the large-$x$ spectators and likely to be described by an emergent effective theory. A natural question is then, since they are both quantum degrees of freedom in the original QFT, how the small-$x$ and large-$x$ partons entangle with each other and how the entanglement is related to experiments.

The entanglement entropy in diffractive $ep$ or $pp$ scattering was first noted in the strong coupling regime using a
 holographic string analysis~\cite{Stoffers:2012mn} (called quantum entropy there), and in weak coupling using the evolution equation based analysis~\cite{Kharzeev:2017qzs}.
 In both cases, the entanglement entropy was found to be extensive in the rapidity, an observation made since by
 many others in perturbative QCD~\cite{Armesto:2019mna,Dvali:2021ooc}. The large entanglement entropy stored in hadrons
 and nuclei, may explain the prompt entropies released in current hadron colliders, in the form of large particle multiplicities
 \cite{Qian:2014rda,Qian:2015boa,Shuryak:2017phz,Liu:2022ohy}. For completeness, we note that an
 entropy composed of the multiplicities of the produced gluons  in the context of saturation models was
 also discussed in~\cite{Kutak:2011rb}.

Recently, we have  used 2D QCD in the large $N_c$ limit, to analyze quantum entanglement in confined meson
 states~\cite{Liu:2022ohy}. The entanglement entropy in parton-x was found to be also extensive in the meson rapidity, but with a
 central-charge-like  played by the cumulative quark PDF,  and hence dependent on the fraction of parton-x measured.
 Much like in 4D, the entanglement entropy exhibits an asymptotic expansion that is similar to the one observed
 for meson-meson scattering in the Regge limit. Most notably, the entanglement entropy per unit rapidity  in a 2D nucleus (a sum of
 longitudinal mesons on the light front) was found to be at the bound set for quantum information flow~\cite{le1967proceedings,Bekenstein:1981zz}.

The purpose of this work is to extend some of our recent 2D QCD observations~\cite{Liu:2022ohy}, to 4D QCD where non-trivial
rapidity divergences are the lore at weak coupling, and closely related to the Regge behavior of scattering amplitudes.
We will mostly focus on the entanglement entropy  in a quarkonium state at next to leading order in the weak coupling
$\alpha_s$, and show that the rapidity divergences are at the origin of double logarithms in rapidity. This observation
is readily extended to higher orders, and re-summed through an evolution equation in the large $N_c$ limit. An
analysis of the proton wave function with a similar motivation was recently discussed in~\cite{Dumitru:2022tud}.

The outline of the paper is as follows:  in section~\ref{sec_QQG} we outline the construction of the reduced density matrix for a quarkonium state in leading order in $\alpha_s$. In section~\ref{sec_REDUCED},
we derive the reduced density matrix for the soft gluon, by tracing over the quark sources. The ensuing entanglement
entropy receives contributions from both the real and virtual parts of the wavefunction, but the latter generates
double logarithms in rapidity as naturally expected. In section~\ref{sec_EVOLUTION}, we show how to resum
the leading rapidity logarithms in the entanglement entropy. In the large $N_c$ limit and weak coupling, the density matrices with
and without overall longitudinal momentum conservation, are shown to obey BK-like evolution equations. The
ensuing entanglement entropy is found to saturate the Kolmogorov-Sinai bound of 1, for the case after partial tracing which imposing momentum conservation, but otherwise linear in the rapidity.  The evolution equation can be solved explicitly  for non-conformal QCD in 3D. The rate of growth of the entanglement entropy is shown to vanish
at large rapidities, a consequence of the shrinking of the transverse phase space. In section~\ref{sec_ADS},
we show that the evolution of the trace of the density matrix obeys a standard BFKL evolution equation, which
can be mapped on an evolution in  an emergent AdS$_3$ space. In section~\ref{sec_STRING}, we extend this
observation to the strong coupling regime, where the evolution is captured by the tachyonic mode of a quantum
string in AdS$_5$, where the string bits are dual to the wee partons. Our conclusions are in section~\ref{sec_CONCLUSION}.

\section{Quarkonium wave function}~\label{sec_QQG}

In this section we study the rapidity-space entanglement of the light front wave function (LFWF) for a pair of a heavy quark and anti-quark. The same system has been investigated in the literature,  to derive the non-linear rapidity evolution equation of the dipole generating functional~\cite{Mueller:1993rr,Mueller:1994gb}. It has also been revisited  recently,  to provide a concrete example for the formulation of LFWF amplitudes~\cite{Ji:2021znw}.

Unlike the 2D case where there is no rapidity divergences in the light front (LF) quantization, in 4D gauge theory there are non-trivial rapidity divergences (RD). They are closely related to the Regge limit of gauge theory. Still, they are not totally understood from first principles, even in perturbation theory (PT).

On the LF, the soft gluon wave function of  a $\bar QQ$ pair, to leading order is of the form~\cite{Mueller:1993rr,Kovchegov:2012mbw}
\begin{align}\label{eq:oneloop}
|\bar Q Q\rangle =|\bar Q Q\rangle^{0}+|\bar Q Q\rangle ^{(1)}+|\bar Q Q g\rangle \ ,
\end{align}
Here
\begin{align}
\label{PT1}
|\bar QQ\rangle ^{(0)}=\frac{1}{\sqrt{\Lambda^-}\sqrt{V_\perp}}\sum_{z,\vec{k}_\perp}\frac{1}{\sqrt{(2\pi)^2}}\Psi^{(0)}_{\sigma,\sigma'}(z,\vec{k}_\perp)|z,\sigma,\vec{k}_\perp\rangle_Q \bigotimes|\bar z,\sigma',-\vec{k}_\perp \rangle_{\bar Q} \ ,
\end{align}
is the leading order quarkonium wave function with initial profile $\Psi^{(0)}_{\sigma,\sigma'}(z,\vec{k}_\perp)$.
The free-Fock basis is normalized as $\langle z|z^\prime\rangle=\delta_{z,z^\prime}$,
and the momentum fraction relates to the discrete LF label $n$ through the relation
\begin{align}
z=\frac{2n+1}{2\Lambda^-} \ .
\end{align}
Comparing with the standard notation in the literature, we have absorbed a factor of ${1}/{\sqrt{z(1-z)}}$ into the definition of the LFWF, which will generates precisely the Lorentz invariant LF phase space measure
\begin{align}
\int_{z_{\rm min}}^1 \frac{dz}{z} \ ,
\end{align}
after squaring. Again, the natural rapidity cutoff $z_{\rm min}$ is given by the inverse of $\Lambda^-$.  We assume that the leading WF $\Psi^{(0)}$ is supported well away from $z=0$, for instance it can be a pair of free quarks with fixed momentum fractions $z=z_0\sim \frac{1}{2}$, and zero transverse momenta $\vec{k}_\perp=0$. More specifically,   $\Psi^{(0)}=\delta_{z,z_0}\delta_{\vec{k}_\perp,0}$, which is the case considered before when formulating the LFWF amplitudes.

The most interesting contribution in (\ref{eq:oneloop}) is the soft-gluon contribution. Using the standard rule in LFPT, one can show that the leading soft gluon contribution to the LFWF is
\begin{align}
|\bar Q Q g\rangle=\frac{1}{\Lambda^-V_\perp}\sum_{z,x,k_\perp,k_g,\epsilon,a}\Psi^a(x,z,\vec{k}_\perp,\vec{k}_{g,\perp},\vec{\epsilon})|z,\vec{k}_\perp,\sigma\rangle_Q|\bar z-x,-\vec{k}_\perp-\vec{k}_{g,\perp},\sigma'\rangle_{\bar Q}|x,\vec{k}_{g,\perp},\vec{\epsilon},a\rangle_g \ ,
\end{align}
where one has
\begin{align}\label{eq:oneloopmom}
\Psi^a(x,z,\vec{k}_\perp,\vec{k}_{g,\perp},\vec{\epsilon})=\frac{2gt^a}{\sqrt{x}}\frac{\vec{\epsilon}^{\star}\cdot \vec{k}_{g,\perp}}{k^2_{g,\perp}}\bigg(\Psi^{(0)}_{\sigma,\sigma'}(z,\vec{k}_\perp+\vec{k}_{g,\perp})-\Psi^{(0)}_{\sigma,\sigma'}(z,\vec{k}_\perp)\bigg) \ .
\end{align}
Notice that the above equation applies to any transverse dimensions, in particular, for $D_\perp=2$ the wave function can be written in coordinate space as
\bea
\Psi^a(x,z,\vec{b}_{1\perp}, \vec{b}_{2\perp})&=&\int \frac{d^2k_\perp d^2k_{g,\perp}}{(2\pi)^4}e^{i\vec{k}_{\perp} \cdot \vec{b}_{10\perp}+i\vec{k}_{g,\perp}\cdot \vec{b}_{20\perp}}\Psi^a(x,z,\vec{k}_\perp,\vec{k}_{g,\perp},\vec{\epsilon})\nonumber \\
&=&\frac{ig t^a}{\pi \sqrt{x}}\Psi^{0}(z,\vec{b}_{1\perp})_{\sigma,\sigma'}\vec{\epsilon}^{\star}\cdot\bigg(\frac{\vec{b}_{21}}{|b_{21}|^2}-\frac{\vec{b}_{20}}{|b_{20}|^2}\bigg) \ ,
\eea
with $\vec{b}_{21}=\vec{b}_{2\perp}-\vec{b}_{1\perp}$, $\vec{b}_{10}=\vec{b}_{1\perp}-\vec{b}_{0\perp}$ and $\vec{b}_{20}=\vec{b}_{2\perp}-\vec{b}_{0\perp}$.  This is  the real part of the wave function. The norm of this state is given by
\begin{align}
\langle \bar Q Q g| \bar Q Q q\rangle=\frac{2\alpha_sC_F}{\pi}\int d^2b_\perp \ln b^2_\perp \mu^2 \times \int^{x_0}_{x_{\rm min}} \frac{dx}{x}\times \int dz \sum_{\sigma,\sigma'} |\Psi^{0}(z,\vec{b}_\perp)_{\sigma,\sigma'}|^2 \ .
\end{align}
There is a UV divergence which is regularized by the cutoff $\mu$, and the rapidity divergence is cutoff by $x_{\rm min}=\frac{1}{2\Lambda^-}$.   On the other hand, the virtual part $|\bar Q Q\rangle ^{(1)}$ can be written with the following wave function
\begin{align}
\Psi^{(1)}(z,\vec{b}_\perp,\sigma,\sigma')=-\frac{\alpha_sC_F}{\pi}\ln b^2_\perp \mu^2 \times \int^{x_0}_{x_{\rm min}} \frac{dx}{x}\times  \Psi^{0}(z,\vec{b}_\perp,\sigma,\sigma') \ ,
\end{align}
With this in mind, it is clear that when squaring (\ref{eq:oneloop}), the real and virtual parts cancel, in agreement with perturbative unitarity.

\section{Soft-gluon entanglement  entropy in quarkonium}~\label{sec_REDUCED}

In our previous work in 2D QCD~\cite{Liu:2022ohy}, we found that for a meson state quantized in the discrete LF quantization, the entanglement entropy in rapidity space is finite in the ultra-violet, but contains a logarithmic divergent term in the effective box size $\Lambda^-=P^+L^-$, which is identified with the meson rapidity. In this section, we investigate the rapidity space entanglement for the quarkonium or $\bar QQ$ system in 4D pertubrative QCD. We show that the non-trivial rapidity divergence (or rapidity transcendentality) leads to an enhanced divergence in $\ln \Lambda^-$. At order $\alpha_s$ the leading contribution is enhanced  by $\ln^2 \Lambda^-$,  and at order $\alpha_s^k$ it is enhanced by $(\ln \Lambda^{-})^{1+k} $.  We first consider the order $\alpha_s$ case, and then generalize  to order $\alpha_s^k$.

\subsection{Leading order in $\alpha_s$}

With the one-loop soft gluon part of the wave function at hand, we can now perform partial tracing,
 to obtain the reduced density matrix. We trace over the quark longitudinal and transverse contributions,
 leaving the gluon longitudinal contribution in the final state untraced,
\bea
\hat \rho=&& |0\rangle \langle0|\bigg(1-\frac{2\alpha_sC_F}{\pi}\int d^2b_\perp \ln b^2_\perp \mu^2 \times \int^{x_0}_{x_{\rm min}} \frac{dx}{x}\times \int dz \sum_{\sigma,\sigma'} |\Psi^{0}(z,\vec{b}_\perp)_{\sigma,\sigma'}|^2\bigg) \nonumber \\
&&+\frac{1}{\Lambda^-}\sum_{x<x_0}\frac{1}{x}|x\rangle_g\langle x|_g\times\bigg( \frac{2\alpha_sC_F}{\pi}\times \int d^2b_\perp \ln b^2_\perp \mu^2 \times \int dz \sum_{\sigma,\sigma'} |\Psi^{0}(z,\vec{b}_\perp)_{\sigma,\sigma'}|^2\bigg)
\eea
Clearly, the off-diagonal term in the gluon longitudinal momentum, vanishes in the trace to this order.

With the reduced density matrix, we now compute the entanglement entropy. The virtual or vacuum part leads to the result
\bea
S_{\rm virtual}=&&-\bigg(1-\frac{2\alpha_sC_F}{\pi}\int d^2b_\perp \ln b^2_\perp \mu^2 \times \int^{x_0}_{x_{\rm min}} \frac{dx}{x}\times \int dz \sum_{\sigma,\sigma'} |\Psi^{0}(z,\vec{b}_\perp)_{\sigma,\sigma'}|^2\bigg)\nonumber \\
&&\times \ln \bigg(1-\frac{2\alpha_sC_F}{\pi}\int d^2b_\perp \ln b^2_\perp \mu^2 \times \int^{x_0}_{x_{\rm min}} \frac{dx}{x}\times \int dz \sum_{\sigma,\sigma'} |\Psi^{0}(z,\vec{b}_\perp)_{\sigma,\sigma'}|^2\bigg)\nonumber \\
=&&\frac{2\alpha_sC_F}{\pi}\int d^2b_\perp \ln b^2_\perp \mu^2 |\Psi^{(0)}(b_\perp)|^2\times \int^{x_0}_{x_{\rm min}} \frac{dx}{x} +{\cal O}(\alpha_s^2 )\ .
\eea
where $|\Psi^{(0)}(b_\perp)|^2 \equiv \int dz \sum_{\sigma,\sigma'}|\Psi^{(0)}(z,b_\perp)_{\sigma,\sigma'}|^2$ is the initial dipole wave function integrated over $z$, and the rapidity divergence (RD) contribution reads
\begin{align}
\int^{x_0}_{x_{\rm min}} \frac{dx}{x}\rightarrow \sum_{n=n_0}^{\Lambda^--1/2}\frac{1}{n+\frac{1}{2}} =\ln \Lambda^- x_0 +C \ .
\end{align}
On the other hand, what is more interesting is the contribution from the real emission.  In terms of
\begin{align}
p_0=\frac{2\alpha_sC_F}{\pi}\int d^2b_\perp \ln b^2_\perp \mu^2 |\Psi^{(0)}(b_\perp)|^2 \ ,
\end{align}
the real contribution to the entanglement entropy is
\bea
S_{\rm real}=\ln \Lambda^- \times p_0\times \int_{x_{\rm min}}^{x_0} \frac{dx}{x}-\int_{x_{\rm min}}^{x_0} \frac{dx}{x} p_0\ln \frac{p_0}{x}
 \rightarrow \frac{1}{2} \ln^2 \Lambda^- p_0+\ln \Lambda^- p_0\ln \frac{x_0}{p_0} +C  \nonumber\\
\eea
The entanglement entropy is now dominated by the double-logarithmic divergent term
\begin{align}
S_{\rm real}+S_{\rm virtual}=\frac{1}{2} \ln^2 \Lambda^- p_0+\ln \Lambda^- p_0\ln\frac{x_0}{p_0} +p_0\ln \Lambda^- x_0+C \ .
\end{align}
It is the RD that leads to the enhancement of the logarithmic dependency on $\Lambda^-$. It is also divergent in UV.

\subsection{Higher  order in $\alpha_s$}
We now generalize the result to all orders, to extract  the leading $\alpha_s \ln \Lambda^-$ contribution. The above suggests the $\alpha_s^k \ln (\Lambda^-)^{k+1}$ as the general pattern, which can be shown as follows. First, to obtain the leading $\ln \Lambda^-$ contribution, it is much better if all the gluons are non-traced, namely, more soft gluons in the final state, more logarithms in $\Lambda$. Therefore we do not perform partial tracing in gluons. However, the quarks should be traced out, which leads to the structure of the entanglement density matrix in the following form
\begin{align}
\label{SCH}
\rho=\sum_{n} p_n\rho_n\ ,
\end{align}
where $p_n$ is the total probability of finding $n$-soft gluons,  and the $\rho_n$ is an effective reduced density matrix with $n$-soft gluon on the left and right.  Notice that the off-diagonal terms in particle numbers simply vanish,  thanks to the total trace in the transverse momentum or color.  From (\ref{SCH}) the entanglement entropy for $\rho$ can be found to be
\begin{align}
S=-\sum_n p_n \ln p_n +\sum_{n} p_n S_n \ ,
\end{align}
where $S_n=-{\rm tr}\rho_n \ln \rho_n$ is the entanglement entropy of the reduced density matrix in the $n$-particle sector.  For purely real soft emissions, the probability of the state is proportional to
\begin{align}
p_n \sim \alpha_s^n \int_{x_{\rm min}}^{x_0} \frac{dx_1}{x_1}\int_{x_1}^{x_0}\frac{dx_2}{x_2}.....\int_{x_{n-1}}^{x_0}\frac{dx_n}{x_n}\propto \alpha_s^n\ln^{n} \Lambda^- \ ,
\end{align}
and adding virtual corrections will correct it in powers of $\alpha_s \ln \Lambda^-$ to leading log accuracy. On the other hand, for the purely real soft emissions one can show that
\begin{align}
\rho_{n,real}\propto \sum_{x_1\gg x_2...\gg x_n;x_1\gg x_2'....\gg x_n'} \frac{1}{x_1\sqrt{x_2x_2'...x_nx_n'}}|x_1\rangle..|x_n\rangle \langle x_1|....\langle x_n'| \ ,
\end{align}
which involves $\Lambda^-$ pure states and consequently $S_n \propto \ln \Lambda^-$. When multiplied by  $\alpha_s^n$ in $p_n$, this generates the desired $\alpha_s^n \ln^{n+1} \Lambda^-$ contribution. With the inclusion of virtual emissions, one expects that the modification will be in order of $\ln \Lambda^- \alpha_s$ as for the $p_n$, therefore will not change the leading logarithmic pattern $\alpha_s^k \ln^{k+1} \Lambda^-$.  Moreover, the leading $\alpha^k \ln^{k} \Lambda^-$ and the next to leading $\alpha^k \ln^{k+1} \Lambda^-$ contributions to the entanglement entropy, can be obtained from the same set of soft gluon LFWFs,  that leads to the leading norm-square of the state $p_n$. This will be exploited further below.

\section{BK-like evolution of the entanglement density}~\label{sec_EVOLUTION}

We now derive an evolution equation for the reduced density matrix in Eq.~(\ref{SCH}) based on the evolution equation for the dipole generating functional in the large $N_c$ limit. We first review the derivation of the evolution equation of the dipole's wave function. Since we trace out all the transverse degrees of freedom,  and leave only the momentum fractions un-traced, it is more convenient to consider the square of the wave function.

\subsection{Evolution of the quarkonium wavefunction}

\begin{figure}[!htb]
\includegraphics[height=5cm]{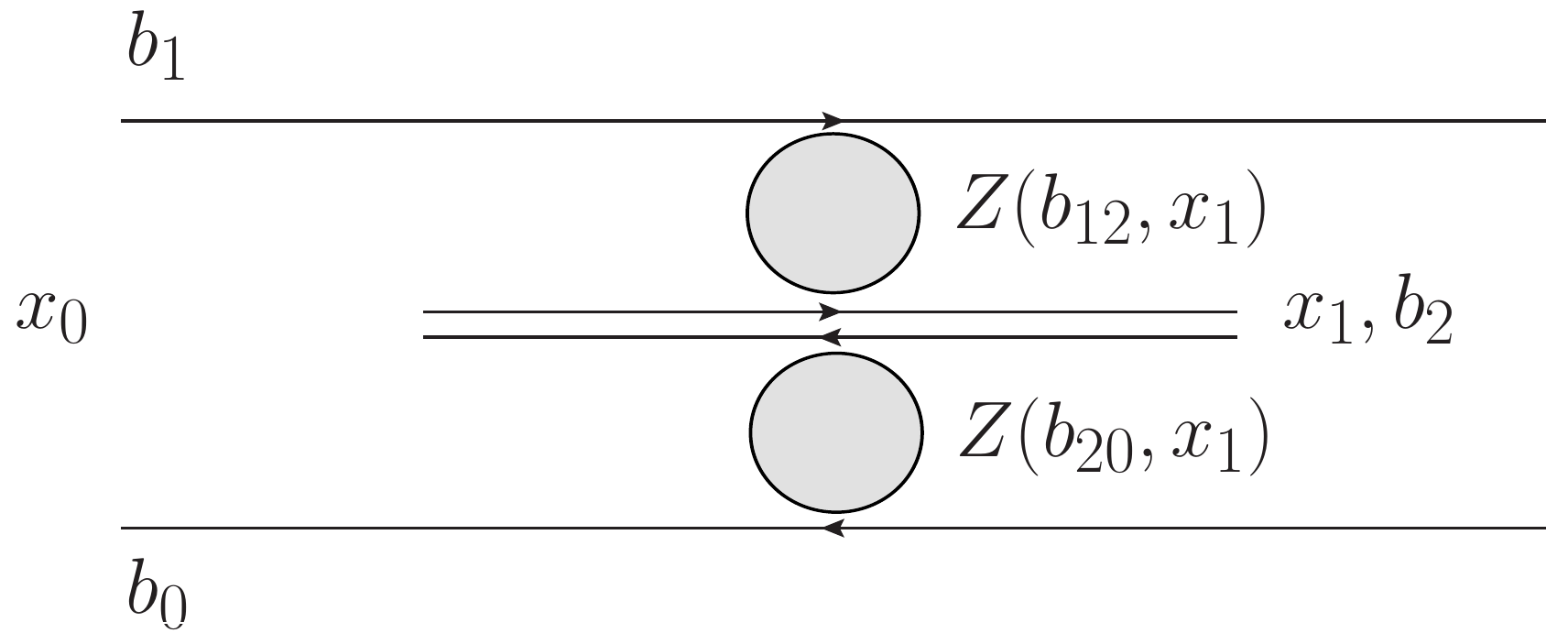}
 \caption{Depiction of the second term of the evolution equation (\ref{eq:evoZ}). The first emitted soft gluon at rapidity $x_1$ and transverse position $b_2$,  splits the original dipole into two dipoles with transverse separations $b_{12}$ and $b_{20}$, within which subsequent soft emissions occur independently. Virtual emissions before the first real gluon with rapidities between $x_0$ and $x_1$, contribute to the ``soft factor'' $S(b_{10},\frac{x_0}{x_1})$. When combined with the purely virtual contribution or the first term in~(\ref{eq:evoZ}), unitarity is restored.  }
  \label{fig:evoZ}
\end{figure}

The perturbative process that leads to the leading RD in the dipole's wave function, is by now standard~\cite{Mueller:1993rr,Mueller:1994gb,Kovchegov:2012mbw}. We briefly recall it here for completeness. The soft gluons are emitted consecutively with strong ordering in their momentum fractions $x_1\gg x_2 \gg x_3... \gg x_n$. The emission of the first soft gluon at transverse position $b_2$,  leads the following factor for the wave function square
\begin{align}
|\Psi^{(1)}|^2=\int_{x_{\rm min}}^{x_0} \frac{dx_1}{x_1}\frac{\alpha_sC_F}{\pi^2} \int d^2 b_2 \frac{b_{10}^2}{b_{21}^2b_{20}^2}|\Psi^{(0)}(b_{10})|^2 \ .
\end{align}
Following  the emission of the first soft gluon,  the  original dipole with transverse size $b_{10}$, splits into  two dipoles with sized $b_{12}$ and $b_{20}$.  The subsequent emissions of  these two dipoles is independent
\begin{align}
|\Psi^{(2)}|^2=(\frac{\alpha_sC_F}{\pi^2})^2 \int_{x_{\rm min}}^{x_0}\frac{dx_1}{x_1}\int_{x_{\rm min}}^{x_1} \frac{dx_2}{x_2}\int d^2b_3\int d^2b_2 \frac{b_{10}^2}{b_{21}^2b_{20}^2}\bigg(\frac{b_{12}^2}{b_{31}^2b_{21}^2}+\frac{b_{20}^2}{b_{23}^2b_{30}^2}\bigg)|\Psi^{(0)}(b_{10})|^2 \ .
\end{align}
The process clearly is repetitive, with  more and more soft gluons emitted. To generalize to all orders, we define the generating functional $Z(b_{10},x_0,u(x,z))$,  such that
\begin{align}
\frac{1}{n!} \frac{\delta^n\, Z(b_{10},x_0,u(b,z))|\Psi^{(0)}(b_{10})|^2}{\delta u(b_2,z_1)\delta u(b_4,z_2)....\delta u(b_{n+1},x_n)}
 = |\Psi^{(n)}(b_{10},x_0;b_2,x_1;b_3,z_2;...b_{n+1},x_n)|^2 \ ,
\end{align}
Here $$|\Psi^{(n)}(b_{10},x_0;b_2,z_1;b_3,z_2;...b_{n+1},z_n)|^2$$ refers to the square of the wave function with $n$ soft gluons, with  transverse positions $b_2$,..$b_{n+1}$ and momentum fractions $x_1>x_2>...x_n$.  This emission cascade process,  satisfies the equation
\begin{align}\label{eq:evoZ}
&Z(b_{10},x_0,x_{\rm min},u)=S(b_{10},\frac{x_0}{x_{\rm min}})\nonumber \\
&+\frac{\alpha_s C_F}{\pi^2} \int_{x_{\rm min}}^{x_0} \frac{dx_1}{x_1} S(b_{10},\frac{x_0}{x_1}) \int db_2^2\frac{b_{10}^2}{b_{12}^2b_{20}^2}u(x_1,b_2)Z(b_{12},x_1,x_{\rm min},u)Z(b_{20},x_1,x_{\rm min},u) \ .
\end{align}
Here we have included the contribution from virtual emissions that give rise to the Sudakov-like suppression factor
\begin{align}
\label{SUDAKHOV}
S\bigg(b_{10},\frac{x_0}{x_1}\bigg)=\exp \bigg[-\frac{2\alpha_s C_F}{\pi}\ln b_{10}^2\mu^2 \ln \frac{x_0}{x_1}\bigg] \ .
\end{align}
It resums the virtual gluon emissions, prior to the emission of the first real gluon inside the first dipole with size $b_{10}$. The constraint that the rapidity of the virtual emission should be between $x_0$ and the next soft gluon $x_1$,  leads to the  $\ln \frac{x_0}{x_1}$ contribution. When  $u\equiv 1$, unitarity requires $Z=1$, which is manifest in the above equation. The evolution equation is depicted in Fig.~\ref{fig:evoZ}.

\subsection{Evolution of the density matrix}
Given the above, let's consider the reduced density matrix. Clearly, we need all soft gluons in the final states being untraced in rapidity. Since they will be traced in color and in transverse positions, the reduced density should have the same spatial factor as for the squared wave function, except that the momentum fractions on the left-ket and right-bra, are different. Namely, we can have $x_1\gg x_2\gg x_3.....\gg x_n$ for the ket and $x_1'\gg x_2'...\gg x_n'$ for the bra. However, since the total momentum fraction has to be the same, the largest momentum fractions are approximately equal, namely, $x_1 \sim x_1'$. Also, since in the large $N_c$ limit there is no crossing, we still have the same diagrammatic depiction as in the case of $Z$. Moreover, to get the dominant contribution in $\alpha \ln \Lambda^-$ we expect that the orderings in the bras and kets are one-to-one. Namely, if in the first dipole  $b_{12}$, the $k$ soft gluons in kets are labeled by $x_{\pi(1)},...x_{\pi(k)}$ where $\pi$ is a permutation of $1,2..n$, then in the bras the soft gluons must be $x'_{\pi(1)},...x'_{\pi(k)}$. And similarly for the remaining $n-k$ soft gluons in the dipole $b_{20}$. More precisely, if the wave function reads
\begin{align}
&|\bar Q Q(b_{10};x_0,x_{\rm min})\rangle\nonumber \\
&=\sum_n \sum_{x_0\gg x_1\gg x_2\gg x_3..\gg x_n \gg x_{\rm min}}f_n(b_{10};x_1,b_1;x_2,b_2;....x_n,b_n)|x_1,b_1\rangle|x_2,b_2\rangle|x_3,b_3\rangle....|x_n,b_n\rangle |\bar Q\rangle |Q\rangle \ ,
\end{align}
then with the following {\it definitions} of the reduced density matrices
\begin{align}
\label{RHO1DEF}
&\rho_1(b_{10};x_0,x_0',x_{\rm min})=\sum_n \int d^2b_1..d^2b_n \nonumber \\
&f_n(b_{10};x_1,b_1;x_2,b_2;..x_n,b_n)f^{\dagger}_{n}(b_{10};x_1',b_1;x_2',b_2;....x_n',b_n)|x_1\rangle|x_2\rangle..|x_n\rangle\langle x_1'|\langle x_2'|...\langle x_n'| \ ,
\end{align}
and
\begin{align}
&\rho(b_{10};x_0,x_{\rm min})=\sum_n \int d^2b_1..d^2b_n \nonumber \\
&f_n(b_{10};x_1,b_1;x_2,b_2;..x_n,b_n)f^{\dagger}_{n}(b_{10};x_1,b_1;x_2',b_2;....x_n',b_n)|x_1\rangle|x_2\rangle..|x_n\rangle\langle x_1|\langle x_2'|...\langle x_n'| \ ,
\end{align}
the locking of the orderings in the bra and ket is automatic in the large $N_c$ limit.  Clearly, this indicates that the soft gluon emissions in the two subsequent dipoles,  can be separated from $x_1,x_1'$ and form two density matrices $\rho_1(b_{12},x_1,x_{\rm min})\otimes\rho_{1}(b_{20},x_1,x_{\rm min})$, namely
\bea
\label{RHO1}
&&\rho_1(b_{10},x_0,x_0^\prime,x_{\rm min},u)=S^{\frac{1}{2}}\bigg(b_{10},\frac{x_0x_0'}{x^2_{\rm min}}\bigg)|0\rangle \langle 0|\nonumber \\
&&+\frac{\alpha_s C_F}{\pi^2} \int db_2^2\frac{b_{10}^2}{b_{12}^2b_{20}^2} \sum_{x_{\rm min}\le x_1,x_1' \le x_0}S^{\frac{1}{2}}\bigg(b_{10},\frac{x_0x_0^\prime}{x_1x_1^\prime}\bigg)\frac{|x_1\rangle\langle x_1^\prime|}{\sqrt{x_1x_1^\prime}}
\otimes \rho_1(b_{12},x_1,x_1^\prime,x_{\rm min})\otimes \rho_1(b_{20},x_1,x_1^\prime,x_{\rm min}) \ .\nonumber\\
\eea
And the $\rho$ can be obtained from $\rho_1$ by simply tracing the hardest gluon
\begin{align}
\label{RHO}
&\rho(b_{10},x_0,x_{\rm min},u)=S\bigg(b_{10},\frac{x_0}{x_{\rm min}}\bigg)|0\rangle \langle 0|\nonumber \\
&+\frac{\alpha_s C_F}{\pi^2} \int db_2^2\frac{b_{10}^2}{b_{12}^2b_{20}^2} \sum_{x_{\rm min}\le x_1 \le x_0}S\bigg(b_{10},\frac{x_0}{x_1}\bigg)\frac{ |x_1\rangle\langle x_1|}{x_1}\otimes \rho_1(b_{12},x_1,x_{\rm min})\otimes \rho_1(b_{20},x_1,x_{\rm min}) \ .
\end{align}
with $\rho_1(b,x_1,x_{\rm min})\equiv \rho_1(b,x_1,x_1,x_{\rm min})$.  We note that to order $\alpha_s$, the earlier perturbative result can be recovered by expanding the above equations.
\begin{figure}[!h]
\includegraphics[height=5cm]{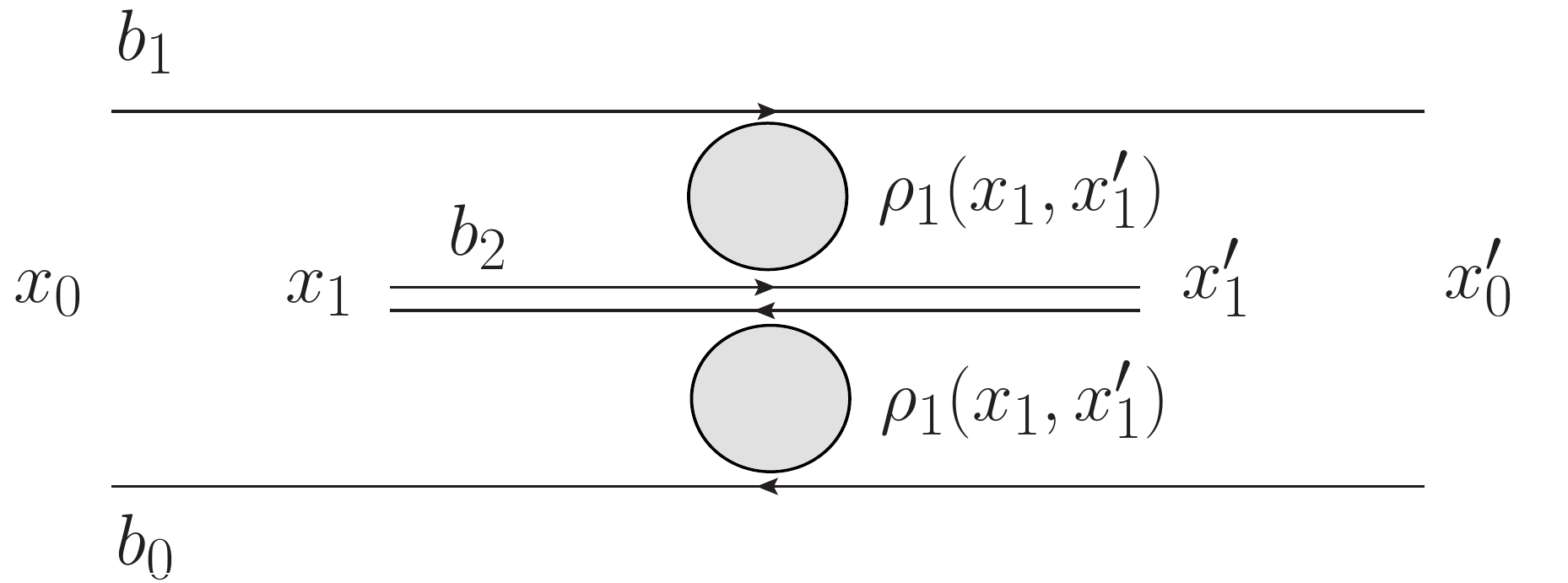}
 \caption{Depiction of the second term of the evolution equation~(\ref{RHO1}). The first emitted soft gluon at rapidity $x_1$ and transverse position $b_2$ splits the original dipole into two dipoles,  with transverse separations $b_{12}$ and $b_{20}$, within which subsequent soft emissions forms the corresponding reduced density matrices $\rho_1(b_{12},x_1,x_1')$ and $\rho_1(b_{20},x_1,x_1')$. Virtual emissions before the first real gluon at left, with rapidities between $x_0$ and $x_1$,   contribute to the ``soft factor'' $S^{\frac{1}{2}}(b_{10},\frac{x_0}{x_1})$, and similarly for $x_0'$,$x_1'$ at right. This process should be regarded as the off-diagonal version of the evolution equation~(\ref{eq:evoZ}).   }
  \label{fig:evorho}
\end{figure}

Here we should note that both $\rho(b_{10},,x_0,x_{\rm min})$ and $\rho_1(b_{10},x_0,x_{\rm min})$ are reduced density matrices. $\rho_1$ can be viewed as the reduced density matrix of the soft gluon wave function,  after tracing over all the degrees of freedom except the rapidity, while $\rho$ can be obtained from $\rho_1$ by further tracing out the rapidity of the in-out $\bar QQ$ pair, which amounts to imposing momentum conservation. (\ref{RHO}-\ref{RHO1}) are the  first major results of this paper.

Unfortunately, due to the nature of the kernel in 4D, the structure of the reduced density matrix is complicated for large $n$.  Below we investigate their entanglement entropies by simplifying the transverse degrees of freedom. We will also show that the partial tracing  of (\ref{RHO1}) describes the rapidity evolution of the one-body density matrix, whose eigenvalues and von Neumann
entropy are measurable in DIS and hadron-hadron scattering in the diffractive regime. In this spirit, the reduced two-body density  following from a pertinent partial tracing of (\ref{RHO1}),
may account for the multiplicities in $pp$ scattering at the  LHC,  with two fixed rapidity gaps.

\subsection{Entanglement entropy in 1+1 reduction}
To simplify the analysis of (\ref{RHO}-\ref{RHO1}), we first consider the case where soft emission is solely longitudinal.
This amounts to freezing the transverse degrees of freedom, and the resulting evolution equation becomes one dimensional.
Note that this is not 2D QCD which is super-renormalizable and with no dynamical gluons. It is more similar to the evolution equation in 4D QCD where the kernel $\frac{b_{10}^2}{b_{12}^2b_{20}^2}$ only exhibit moderate decay at large $b_2$, but no serious constraint.  With this in mind, the 1+1 version of (\ref{RHO}-\ref{RHO1}), is best analysed by
following Mueller~\cite{Mueller:1994gb}, and by introducing the generating functional with a constant soft gluon current $u$
\begin{align}\label{eq:evo1D}
Z(y,u)=e^{-ay}+aue^{-ay}\int_{0}^y e^{ay_1}dy_1 Z(y_1,u)Z(y_1,u) \  .
\end{align}
We have identified the rapidity of the $\bar QQ$ pair with $y$
\begin{align}
y=\ln \frac{x_0}{x_{\rm min}}=\ln x_0\Lambda^- \ .
\end{align}
and reduced the Sudakhov factor (\ref{SUDAKHOV}) to $e^{-ay}$, by assuming $b$-independence
\bea
\frac{2\alpha_s C_F}{\pi}\ln b_{10}^2\mu^2\rightarrow a
\eea
The above equation can be easily solved by iterating Eq.~(\ref{eq:evo1D})
\begin{align}
Z(y,u)=\sum_{n=0}^{\infty} e^{-ay}u^n\bigg(1-e^{-ay}\bigg)^{n} \ ,
\end{align}
with $Z(y,1)=1$. From the above, we  can read the probability of finding $n$ soft gluons as
\begin{align}
\label{PNX}
p_n=e^{-ay}\bigg(1-e^{-ay}\bigg)^{n}=\frac 1{\bar n+1}\bigg(1-\frac 1{\bar n+1}\bigg)^n
\end{align}
with $\bar n=\sum_{n} np_n=e^{ay}-1$ the mean number. As noted in~\cite{Mueller:1994gb}, we have $\bar n p_n\approx e^{-n/\bar n }$ for large
mean $\bar n$, in agreement with KNO scaling.
\\
\\
\noindent{\bf Density matrices:}
\\
The reduced density matrix for the untraced soft gluons {\it without} overall momentum conservation is
\bea
\hat \rho_1= \sum_{n=0}^{\infty} |\Psi_n\rangle \langle \Psi_n| \ ,
\eea
with the effective state
\bea
|\Psi_n\rangle=\sum_{x_0\gg x_1\gg x_2...\gg x_n \gg x_{\rm min}}\sqrt{a^n n!}\frac{e^{-\frac{ay+a\sum_{i=1}^ny_i}{2}}}{\sqrt{(\Lambda^-)^nx_1..x_n}}|x_1\rangle |x_2\rangle.....|x_n\rangle \ ,
\eea
where $y_i=\ln \frac{x_i}{x_{\rm min}}$. Notice that for each $n$,   $\rho_1$ is already in the diagonal form, therefore it is the full reduced density matrix without assumptions in the ordering of $x_1,x_2,..x_n'$. In contrast, the one {\it with} momentum conservation is
\bea
\rho=\sum_{n}\sum_{x_1}|\Psi_{x_1},n\rangle \langle \Psi_{x_1},n| \ ,
\eea
with the state
\bea
|\Psi_{x_1},n\rangle=\frac{e^{-\frac{ay_1}{2}}}{\sqrt{\Lambda^-x_1}}|x_1\rangle \sum_{x_1\gg x_2...\gg x_n \gg x_{\rm min}}\sqrt{a^n n!}\frac{e^{-\frac{ay+a\sum_{i=1}^ny_i}{2}}}{\sqrt{(\Lambda^-)^{n-1}x_2..x_n}} |x_2\rangle.....|x_n\rangle \ .
\eea
Both reduced density matrices satisfy the BK-like evolution equations (\ref{RHO}-\ref{RHO1})
\\
\\
\noindent{\bf Entanglement entropy from  $\hat\rho_1$:}
\\
The entanglement entropy associated to $\rho_1$, for the soft gluon wave function without the overall momentum
constraint (with  the longitudinal momentum of the $\bar QQ$ untraced), is readily obtained using the von Neumann entropy
and the gluon probabilities (\ref{PNX})
\begin{align}
S(\hat \rho_1)=-\sum_{n=0}^{\infty}p_n\ln p_n=ay-\ln(1-e^{-ay})\bigg(e^{ay}-1\bigg) \rightarrow ay+1+{\cal O}\left(e^{-ay} \right) \ ,
\end{align}
It asymptotes  $ay$ for large rapidity $y\rightarrow \infty$, with $a\sim \alpha_s C_F$.
 This result was noted when analysing DIS scattering at weak coupling in the Regge limit~\cite{Kharzeev:2017qzs},
 and hadron-hadron scattering also in the Regge limit at strong coupling in~\cite{Stoffers:2012mn} (although in the
 latter it was initially identified as a quantum entropy).

In the Regge limit, DIS
 and hadron-hadron scattering are universally described by dipole-dipole scattering~\cite{Mueller:1994gb}. In weak coupling,
 the scattering is dominated by exchange of BFKL pomerons. At strong coupling,
 the scattering is dominated by closed string exchanges. The entanglement entropy controls
 the rise of the low-$x$ gluons in DIS, and the rise of the large-s elastic cross section
in diffractive hadron-hadron scattering.

The eigenvalues $p_n$ of the reduced density matrix $\hat\rho_1$, describe  the wee parton multiplicities
at low-x or large $\sqrt{s}$, that may turn real in an inclusive DIS process, or a diffractive hadronic process
with particle production, and KNO scaling. In this sense, the entanglement content of
the reduced density matrix $\hat\rho_1$, with untraced or fixed $\bar QQ$
longitudinal momenta, is directly accessible to DIS, or  hadron-hadron scattering in the Regge limit.
In Nuclei, an even  larger form of entanglement maybe at work, as we suggested recently~\cite{Liu:2022ohy}.
\\
\\
\noindent{\bf Entanglement entropy from $\hat\rho$:}
\\
The reduced density matrix $\hat\rho$ after tracing the $\bar QQ$ is expected to be more entangled, with
a larger entanglement entropy.  Indeed, the entanglement entropy is now of the form
\begin{align}
\label{SRHO1}
S(\rho)=aye^{-ay}-\sum_{n=1}^{\infty}\int_{0}^y dy_1 \ln \bigg[nae^{-a(y+y_1)-y_1}(1-e^{-ay_1})^{n-1}\bigg] nae^{-a(y+y_1)}(1-e^{-ay_1})^{n-1} \ .
\end{align}
To evaluate (\ref{SRHO1}) we split it into three contributions
\begin{align}
&S(\rho)-aye^{-ay}=\sum_{n=1}^{\infty}\int_{0}^ydy_1 \bigg(y_1+a(y+y_1)\bigg) nae^{-a(y+y_1)}(1-e^{-ay_1})^{n-1} \nonumber \\
&-\sum_{n=0}^{\infty}\ln (na)p_n-\sum_{n=0}^{\infty}\int_{0}^ydy_1\ln(1-e^{-ay_1})n(n-1)ae^{-a(y+y_1)}(1-e^{-ay_1})^{n-1} \ .
\end{align}
The first contribution in (\ref{SRHO1}) can be calculated by summing over $n$
\begin{align}
S_1=&\sum_{n=1}^{\infty}\int_{0}^ydy_1 \bigg(y_1+a(y+y_1)\bigg) nae^{-a(y+y_1)}(1-e^{-ay_1})^{n-1}\nonumber \\
=&\int_{0}^y dy_1 a\bigg[y_1+a(y+y_1)\bigg]e^{-a(y-y_1)}=y(1+2a)-\frac{1+a}{a}+{\cal O}(e^{-ay}) \ ,
\end{align}
and then expanding for large rapidity $y$. Similarly, the third contribution in (\ref{SRHO1})  gives
\begin{align}
S_3=&-\sum_{n=1}^{\infty}\int_{0}^ydy_1\ln(1-e^{-ay_1})n(n-1)ae^{-a(y+y_1)}(1-e^{-ay_1})^{n-1}\nonumber \\
=&-2e^{-ay}\int_{0}^y ady_1 \ln (1-e^{-ay_1})(1-e^{-ay_1})e^{2ay_1} \ .
\end{align}
By using the inequality
\begin{align}
-\ln(1-e^{-ay_1})(1-e^{-ay_1})\le e^{-ay_1}\ ,
\end{align}
we have
\begin{align}
S_3 \le 2(1-e^{-ay})\ .
\end{align}
On the other hand, using the inequality   $-\ln (1-e^{-ay_1}) \ge e^{-ay_1}$, we also have
\begin{align}
S_3 \ge 2(1-e^{-ay})-2aye^{-ay} \ .
\end{align}
Thus,  as $y \rightarrow \infty$ we have
\begin{align}
S_3\rightarrow 2 +{\cal O}(ye^{-ay}) \ .
\end{align}
Finally, the second contribution in (\ref{SRHO1})  can be calculated for large $y$ as
\begin{align}
S_2=-\sum_{n=0}^{\infty}\ln (na)p_n\rightarrow -ay- \ln a-\int_{0}^{\infty} dx \ln x e^{-x}=-ay-\ln a+\gamma_{\rm E} \ .
\end{align}
Given the above, we found that the non-vanishing part of the entanglement entropy reads
\begin{align}
 S(\rho)=y\left(1+a\right)+1+\gamma_E-\frac{1}{a}-\ln a=\ln (x_0 \Lambda^-) (1+a)+1+\gamma_E-\frac{1}{a}-\ln a \ .
\end{align}
with $\gamma_E$ being the Euler constant.

In sum, to any given order in $a$, each soft emission, either real of virtual, is accompanied by a
 $y^n$ divergence in the wavefunction,  at order $a\sim \alpha_s C_F$ and  large rapidity $y$. However,
 the re-summed divergences  to order $ay$ are finite and saturate the Kolmogorov-Sinai bound of 1. The bound was
 noted in~\cite{Liu:2022ohy}. The entanglement entropy remains linear in $y$,  despite the growing
 rapidity divergences with increasing order in $\alpha_s$.

\subsection{Entanglement entropies and saturation in 3D QCD}

DIS or hadron-hadron scattering in 3D QCD which is super-renormalizable, non-conformal and confining,
is  still dominated by soft gluon emissions at weak coupling in the Regge limit.
The BK-like equations for the corresponding density matrices (\ref{RHO}-\ref{RHO1})
still hold. Interestingly, this case can be solved exactly with both longitudinal and transverse
evolution in place.

Indeed, in 3D the BK-like evolution equations can be solved using  the generating functional
\begin{align}\label{eq:BK3D}
Z(b,y,u)=e^{-mb y}+um\int_{0}^{y} dy_1 e^{-mb(y-y_1)}\int_{0}^b db'Z(b-b',y_1,u)Z(b',y_1,u) \ ,
\end{align}
with  $m \equiv \frac{g^2_{1+2}C_F}{4\pi^2}$  the mass scale in 3D.  To derive this equation,  we start with (\ref{eq:oneloopmom}), and write it in coordinate space. Using the identity
\begin{align}
\int_{-\infty}^{\infty} \frac{dk^z}{(2\pi)}\frac{e^{ik^z b}}{k^z}=\frac{i}{\pi}\int_{0}^{\infty} \frac{dk^z}{k^z}\sin(k^zb)=\frac{i}{2}{\rm sign} (b) \ ,
\end{align}
we now have the wave function in the coordinate space, after the emission of a single soft gluon
\bea
\Psi^a(x,z,b_1,b_0,b_2,\epsilon)&=&\int \frac{dk_\perp dk_{g,\perp}}{(2\pi)^2}e^{ik b_{10}+k_{g}b_{20}}\Psi^a(x,z,k,k_g,\epsilon)\nonumber \\
&=&\frac{ig t^a}{2\pi \sqrt{x}}\Psi^{0}(z,b_1)_{\sigma,\sigma'}\epsilon\bigg({\rm sign}(b_{21})-{\rm sign}(b_{20})\bigg) \ .
\eea
This means that the emitted gluon position $b_2$ lies between the mother dipole positions $0$ and $1$, in transverse space and consequently the integral over $b_2$ is IR safe. Given the above, the derivation of the evolution equation is identical to the 4D case which yields (\ref{eq:BK3D}) with $m=\frac{\alpha_s C_F}{\pi}$.

It is not hard to show that the solution of the equation above can be solved explicitly. First, the generating functional can be shown to be
\begin{align}
Z(b,y,u)=e^{-mby}\sum_{n=0}^{\infty}u^n\frac{(mby)^n}{n!}=e^{mby(u-1)} \ .
\end{align}
Indeed, when plugged  into~(\ref{eq:BK3D}),  one can  readily verify that the above is indeed the solution with the condition $Z(b,y,1)=1$.  From these,
 one identifies the probability of finding $n+1$ dipoles with the Poisson gluon emissivities
\begin{align}
p_n=e^{-mby}\frac{(mby)^n}{n!} \ .
\end{align}
The reduced density matrix without the momentum constraint can be shown to be
\begin{align}
\hat\rho_1=\sum_{n} \hat\rho_{1,n} =
\sum_{x_1 \gg x_2... \gg x_n,x_1' \gg x_2'...\gg x_n'} \frac{e^{-mby}(mb)^n}{(\Lambda^-)^n\sqrt{x_1x_2...x_n x_1'x_2'...x_n'}} |x_1...x_n\rangle \langle x_1'..x_n'| \ ,\nonumber\\
\end{align}
from which one reads the most singular part of the x-weighted gluon PDF as
\begin{align}
\label{XGX}
xf_g(x)=mb x^{mb} \ ,
\end{align}
which as expected, is independent of $y$ or RD free. (\ref{XGX}) is the gluonic structure function, which is usually accessible
to DIS kinematics in the Regge limit.

Given $\rho_1$, the corresponding density matrix with the momentum constraint $\hat\rho$ follows from $\hat\rho_1$
by setting $x_1=x_1'$.  The von Neumann entropy for $\hat\rho_1$ is
\begin{align}
\label{SRHO1X}
S(\rho_1)=-\sum_{n}p_n\ln p_n  \ ,
\end{align}
while for $\hat\rho$
\begin{align}
\label{SRHOX}
S(\rho)=-p_0\ln p_0-\sum_{n=1}^{\infty}\int dy_1 \frac{y_1^{n-1}(mb)^ne^{-mby}}{(n-1)!}\ln \bigg( \frac{y_1^{n-1}(mb)^ne^{-mby}}{(n-1)!}e^{-y_1}\bigg) \ ,
\end{align}
At large $b$,  the latter is extensive in  $y$. Indeed, using the Stirling's formula
\begin{align}
\sqrt{2\pi n}\bigg(\frac{n}{e}\bigg)^ne^{\frac{1}{12n+1}}<n!<\sqrt{2\pi n}\bigg(\frac{n}{e}\bigg)^ne^{\frac{1}{12n}} \ ,
\end{align}
we can simplify the integrand in (\ref{SRHOX})
\begin{align}
\label{EXPX}
\sum_{n=1}^{\infty}\frac{(mby)^n}{n!}e^{-mby}n\ln n \rightarrow mby\ln (mby)-mby+\frac{1}{2}\ln mby+\frac{1}{2}\ln 2\pi+\frac{1}{2} \ .
\end{align}
The  first term follows from the peak of the Poisson distribution, but the next contributions
require the full formula. Inserting (\ref{EXPX}) in (\ref{SRHO1X}) we obtain for the entanglement entropies
\bea
S(\rho_1)&\rightarrow & \frac{1}{2}\ln (2\pi e mby) \ ,\nonumber\\
S(\rho) &\rightarrow& y+ \frac{1}{2}\ln (2\pi e mby)+1 \ .
\eea
Unlike 4D QCD, the entanglement entropy for $\hat\rho_1$ in the non-conformal 3D QCD, is not extensive in the rapidity $y$.
The rate of growth of the entropy vanishes at large rapidities, a result which is consistent with the lack of growth of the gluonic
structure function in (\ref{XGX}). This kinematical form of {\it saturation} is a consequence of the specific form of the evolution kernel: 
in 3D QCD, the transverse position $b_2$ of the emitted dipole is forced to be {\it within} the original dipole as in (\ref{eq:BK3D}). Therefore, when there are 
more and more dipoles in the wave function, the average size for these dipoles will become smaller and smaller, and the probability for emitting new dipoles will be 
penalized. This is not the case in 4D where the emission of new dipoles with large sizes is not penalized strongly.

Finally, we note that the entanglement entropy for  $\hat\rho$, is similar to that for $\hat\rho_1$ plus an additional  linear
contribution in $y$. This extensivity in $y$ follows from the extra tracing over the $\bar QQ$ pair, and in-out longitudinal
momentum conservation.


\section{Emergent AdS space from rapidity evolution in QCD}~\label{sec_ADS}

The integral equations for the density matrices in (\ref{RHO}-\ref{RHO1}), reflect on the multi-body content of the
soft emissions in the $\bar QQ$ state. They are part of the contributions in onium-onium (dipole-dipole) scattering,
as originally  discussed by Mueller~\cite{Mueller:1993rr,Mueller:1994gb}. A more general reduced matrix that is off-diagonal
in both the rapidity and the dipole size can be inferred from (\ref{RHO1DEF}), by substituting $b_1\rightarrow b_1^\prime$ in the outgoing
amplitude $f^\dagger$ without tracing over $b_1$. After  performing the partial traces
of this entangled density matrix, we obtain the BFKL evolution for the so-called dipole-dipole Green's function. In the process, we will unravel an emergent AdS structure that will allow
us to bridge the perturbative or partonic contribution for soft gluon emissivities, with the non-perturbative string contribution
using string bits.

\begin{figure}[!h]
\includegraphics[height=5cm]{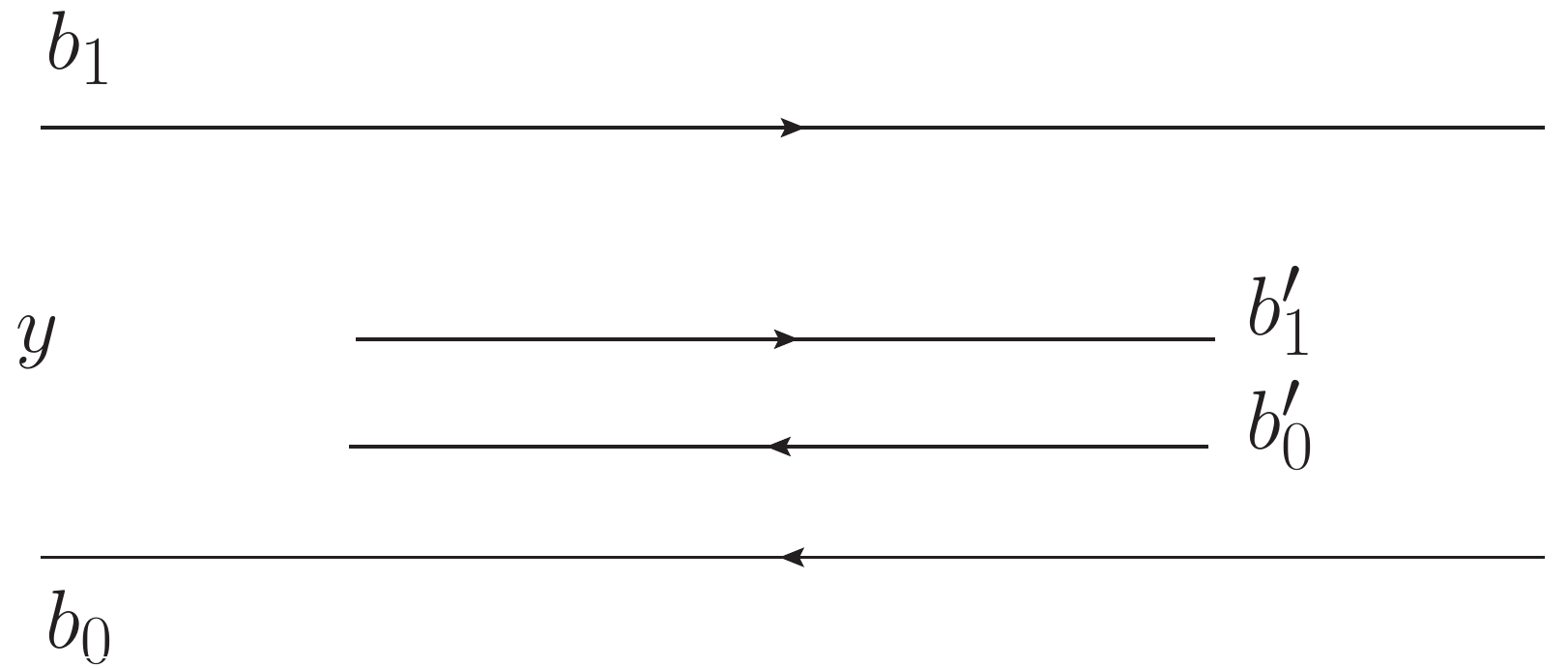}
 \caption{Illustration of the dipole-dipole correlation finction $n(b_{10},b_{10},b,y)$: $b_{10}=b_1-b_0$ is the size of the mother dipole,
 and $b_{10}'=b_1'-b_0'$ the size of the daughter dipol. The impact parameter is given by $b=\frac{b_0'+b_1'}{2}-\frac{b_1+b_0}{2}$ }
  \label{fig:dipolecorre}
\end{figure}

With this in mind, we  introduce the {\it dipole-dipole Green$^\prime$s function} $n(b_{10},b'_{10},b,y)$, defined as the probability of finding a {\it daughter} dipole with size $b'_{10}$ and at impact parameter $b$, within the wave function of the {\it mother} dipole $b_{10}$,
\begin{align}
n(b_{10},b'_{10},b,y)=\frac{\delta \tilde Z(b_{10},b_0,x_0,x_{\rm min},J)}{\delta J(b+b_0,b'_{10}) }|_{J=1} \ ,
\end{align}
where
\begin{align}\label{eq:evotildeZ}
&\tilde Z(b_{10},b_0,x_0,x_{\rm min},J)=S(b_{10},b_0,\frac{x_0}{x_{\rm min}})J(b_0,b_{10})\nonumber \\
&+\frac{\alpha_s C_F}{\pi^2} \int_{x_{\rm min}}^{x_0} \frac{dx_1}{x_1} S(b_{10},\frac{x_0}{x_1}) \int db_2^2\frac{b_{10}^2}{b_{12}^2b_{20}^2}\tilde Z(b_0-\frac{b_{20}}{2},b_{12},x_1,x_{\rm min},J)\tilde Z(b_0+\frac{b_{12}}{2},b_{20},x_1,x_{\rm min},J) \ .
\end{align}
is the generating functional for the dipole wave function squares. The  dipole current $J(b,b_{12})$ refers to a dipole of transverse size  $b_{12}$
centered at $b$ as illustrated in~Fig.~\ref{fig:dipolecorre}. It follows thar $n$ satisfies the evolution equation in rapidity
\begin{align}
\label{TRACEN}
&\partial_y n(b_{10},b'_{10},b,y)=\frac{\alpha_sN_c}{2\pi^2}\int d^2b_2\frac{b_{10}^2}{b_{12}^2b_{20}^2}\nonumber \\
&\bigg[n(b_{12},b'_{10},b-\frac{b_{20}}{2},y)+n(b_{20},b'_{10},b-\frac{b_{21}}{2},y)-n(b_{10},b'_{10},b,y)\bigg] \ .
\end{align}
This equation has an  $SL(2,C)$ symmetry~\cite{Lipatov:1996ts,Kovchegov:2012mbw},  which is more transparent in the holomorphic coordinates $\rho_i=x_i+iy_i$, where $b_i=(x_i,y_i)$ are the transverse positions of the four boundaries $b_0,b_1$, $b_0',b_1'$ of the two dipoles, with $b_{10}=b_1-b_0$ and $b'_{10}=b'_1-b'_0$. In these coordinates, the BFKL equation becomes
\begin{align}
\label{TRACEBFKL}
\partial_y n(\rho_1,\rho_0;\rho_1',\rho_0';y)=\frac{\alpha_sN_c}{2\pi^2}\bigg(H_{GG}+\bar H_{GG}\bigg)n \ .
\end{align}
The BFKL Hamiltonian can be written in terms of the generator of $SL(2,C)$~\cite{Lipatov:1996ts,Nowak:1995ri,Kirschner:1995rn}
\bea
M^z=\rho_1\partial_1+\rho_2\partial_2 \ , \qquad
M^{+}=-\rho_1^2\partial_1-\rho_2^2\partial_2 \ , \qquad
M^{-}=\partial_1+\partial_2 \ ,
\eea
and its Casimir operator
\begin{align}
M^2=(M^z)^2-\frac{1}{2}(M^+M^-+M^-M^+)=-\rho_{12}^2\partial_{1}\partial_2 \ ,
\end{align}
as
\begin{align}
H_{GG}=\sum_{l=0}^{\infty}\bigg(\frac{2l+1}{l(l+1)-M^2}-\frac{2}{l+1}\bigg) \  .
\end{align}
The same applies to the  anti-holomorphic  section. The eigenvalues of the Casimir depend on $n \in Z $ and $\nu \in R$,
\bea
-M^2 E_{n,\nu}=h(h-1)E_{n,\nu} \ ,\qquad -\bar M^2 E_{n,\nu}=\bar h (\bar h-1)E_{n,\nu} \ ,
\eea
with specifically
\bea
h=\frac{1+n}{2}+i\nu \ , \qquad \bar h=\frac{1-n}{2}-i\nu \ .
\eea
and the  eigenvalues
\bea
E_{n,\nu}(\rho_{1a},\rho_{2a})=\bigg(\frac{\rho_{12}}{\rho_{1a}\rho_{2a}}\bigg)^{h}\bigg(\frac{\bar \rho_{12}}{\bar \rho_{1a}\bar \rho_{2a}}\bigg)^{\bar h} \ ,
\eea
In terms of these, it is easy to see that the eigenvalues of $H+\bar H$ are real, equal  and given by the di-gamma function
\begin{align}
\chi(n,\nu)=4{\rm Re}\psi\bigg(\frac{1+|n|}{2}+i\nu\bigg)-4\psi(1) \ .
\end{align}

With this in mind, the solution for  the reduced dipole density $n$ in (\ref{TRACEBFKL}),  can be written in terms of the eigenfunctions
of the BFKL Hamiltonian and its holomorphic section,
\begin{align}
\label{NHYP}
n(\rho_1,\rho_0,\rho_1',\rho_0')=\sum_{n}\int_{-\infty}^{\infty} d\nu C_{n,\nu}e^{y\bar \alpha_s \chi(n,\nu)}\int d^2\rho_a E_{n,\nu}(\rho_{1a},\rho_{0a})E_{n,\nu}(\rho_{1'a},\rho_{2'a}) \ ,
\end{align}
We now note that the integrand in (\ref{NHYP}) can be re-written in terms of hyper-geometrical functions,
using the conformal variable~\cite{Lipatov:1996ts,Kovchegov:2012mbw}
\begin{align}
w=\frac{\rho_{10}\rho_{1'0'}}{\rho_{11'}\rho_{00'}} \ ,
\end{align}
as
\begin{align}
&G_{n\nu}(\rho_1,\rho_0,\rho_1',\rho_0')=c_1x^h \bar x^{\bar h}F[h,h,2h,x]F[\bar h,\bar h,2\bar h,\bar x]\nonumber \\
&+c_2x^{1-h} \bar x^{1-\bar h}F[1-h,1-h,2(1-h),x]F[1-\bar h,1-\bar h,2(1-\bar h),\bar x] \ .
\end{align}
Remarkably,
\begin{align}
x^hF[h,h,2h,x] \ ,
\end{align}
is the scalar propagator   in conformal AdS$_3$ space (identified as  $1_y+1_b+1_z$),  with the invariant AdS$_3$
length $$x \sim \frac{zz'}{(z-z')^2+(b-b')^2} \ .$$ In our case, the transverse $b$ is large with
$x \sim \frac{uu'}{b^2}$.  In this limit,  (\ref{NHYP}) is dominated by the ground state with $n=0$,
\begin{align}
\frac{b_{10}b_{10}'}{b^2}\int d\nu C_{0,\nu}c_1(0,\nu) \bigg(\frac{b_{10}b_{10}'}{b^2}\bigg)^{2i\nu }e^{y\bar \alpha_s \chi(0,\nu)} \ .
\end{align}
with the lowest eigenvalue
\begin{align}
\chi(0,\nu)\sim 4\ln 2-14\zeta_3 \nu^2  \ ,
\end{align}
The result is a  Gaussian integral, which always gives a factor
\begin{align}
\exp \bigg[4\ln 2 \bar \alpha_s y-\frac{\pi}{14\alpha \zeta_3 N_c y}\ln^2 \frac{b^2}{b_{10}b'_{10}} \bigg] \ ,
\end{align}
The pre-factor  depends on the initial condition. For large $b$, one can identify  $(M^z)^2$ with the Laplacian in AdS$_3$,
\bea
z^2(\partial_z^2+\partial_b^2) f\bigg(\frac{zz'}{b^2}\bigg) &\rightarrow& \frac{z^2(z')^2}{b^4}f''\bigg(\frac{zz'}{b^2}\bigg) \ , \\
-\rho^2_{12}\partial_{1}\partial_{2}f\bigg(\frac{\rho_{12}\rho'_{12}}{b^2}\bigg) &\rightarrow& \frac{\rho_{12}^2(\rho')_{12}^2}{b^4}f''\bigg(\frac{\rho_{12}\rho_{12}'}{b^2}\bigg) \ .
\eea
This explains the match in the eigenfunctions.  Note that in AdS$_3$, the eigenvalues for the scalar field are
\begin{align}
i\Delta=\sqrt{-M^2+E} \ , \qquad E=M^2-\Delta^2 \ ,
\end{align}
which match those of the BFKL kernel. Therefore, at  large $b$ there is  an emergent AdS$_3$ structure of the BFKL solution.
The propagator, the Laplacian and the symmetry match.

To explain the pre-factor, requires the initial condition. Specifically, for the initial condition
\begin{align}
n(b_{10},b_{1'0'},b,y=0)=\frac{2\alpha_s C_F}{N_c}\ln^2 \frac{b_{11'}b_{00'}}{b_{10'}b_{01'}} \ ,
\end{align}
we fix the expansion coefficient $C_{n,\nu}$ in (\ref{NHYP}) as
\begin{align}
C_{n,\nu}=\frac{\nu^2+\frac{n^2}{4}}{\bigg(\nu^2+\frac{(n-1)^2}{4}\bigg)\bigg(\nu^2+\frac{(n+1)^2}{4}\bigg)}
\Gamma\bigg(\frac{|n|}{2}+i\nu\bigg)\times \tilde C \  ,
\end{align}
For $n=0$, it behaves as  $i\nu$ for small $\nu$, and is  dominant in the  large $b$,
or large $y$ limits. This leads the additional contribution
\begin{align}
\ln \bigg(\frac{b^2}{b_{10}b'_{10}}\bigg)\frac{1}{y} \ ,
\end{align}
When combined with the factor of  ${1}/{y^{\frac{1}{2}}}$ from the gaussian integral, this yields the expected
rapidity dependent  pre-factor ${1}/{y^{\frac{3}{2}}}$!

\section{String dual description of rapidity evolution in AdS space}~\label{sec_STRING}

The emergence of an AdS space structure that characterizes Mueller dipole evolution~\cite{Mueller:1993rr,Mueller:1994gb} in rapidity in the BFKL limit,
is purely in the perturbative realm of QCD. This is not totally surprising, since the BFKL equation exhibit conformal symmetry. Yet this relationship is useful, as it points to the string dual description in AdS space at strong
coupling~\cite{Rho:1999jm,Janik:2000aj,Polchinski:2002jw,Brower:2006ea,Basar:2012jb,Janik:2013pxa}. Indeed, Feynman wee-parton description (weak coupling)~\cite{Feynman:1969wa},
 is dual to Susskind-Thorn string-bit
description~\cite{Susskind:1993aa,Thorn:1994sw}  (strong coupling), albeit in curved AdS space.  This allows for the extension of
the entanglement entropy calculations at weak coupling~\cite{Kharzeev:2017qzs}, to strong coupling and large $N_c$,
as originally suggested in the form of a quantum entropy in the context of holography~\cite{Stoffers:2012mn}.

In the AdS approach, the holographic direction $z$ is dual to the dipole sizes $x_{10},x'_{10}$, and the impact parameter space can be
of arbitrary $d_{\perp}$ dimensions~\cite{Polchinski:2002jw,Brower:2006ea}. The relevant quantity is the tachyon propagator~\cite{Stoffers:2012mn}
\begin{align}
G(\Delta(j),W)=W^{\Delta}F[\Delta(j),\Delta(j)+\frac{1-d_\perp}{2},2\Delta(j)+1-d_\perp,-4W] \ ,
\end{align}
which can be naturally expressed in the invariant AdS length
\begin{align}
W=\frac{zz'}{(z-z')^2+b^2} \ ,
\end{align}
Here  $\Delta(j)$ is related  to the string tachyon  mass $M_0^2=-(d_\perp+1)/6\alpha^\prime$
($\alpha^\prime\sim 1/\sqrt{\lambda}$ is the
squared string length in units of the AdS radius), through the relation
\begin{align}
\Delta(j)=\frac{d_\perp}{2}+\sqrt{-M_0^2-j-\frac{d_\perp^2}{4}} \ .
\end{align}
The Green-function can be written in this case as
\begin{align}\label{eq:stringpro}
N(T_\perp,zz'/b^2)\sim \int dj e^{jT_\perp} G(\Delta(j),W) \ .
\end{align}

If we  identify $\Delta(j)$ with the weight in the BFKL kernel
\begin{align}
h=\frac{1}{2}+i\nu \ ,
\end{align}
we  find that $d_\perp=1$,  and the variable conjugate to $y=\frac{T_\perp}{{\cal D}}$ as
\begin{align}
{\cal E}={\cal D}j={\cal D}\bigg(-M_0^2-\frac{d_\perp^2}{4}-\nu^2\bigg) \ ,
\end{align}
with ${\cal D}=\alpha^\prime/2$, and therefore
\begin{align}
\label{ALPHAP}
\alpha_{\mathbb P}={\cal D}\left(-M_0^2-\frac{d_\perp^2}{4} \right) =\frac{\alpha^\prime}2 \bigg(\frac{d_\perp+1}{6\alpha^\prime}-\frac{{d_\perp}^2}4\bigg)\ .
\end{align}
with $\alpha_{\mathbb P}$ the Pomeron intercept.
Since $M_{0}^2 \propto -\frac{1}{\alpha^\prime}$ and ${\cal D} \propto \alpha^\prime$, the dominant contribution clearly comes from the
tachyon mass  $-M_0^2$.  Similarly, in the BFKL case the conjugate variable to $y$ reads
\begin{align}
{\cal E}=\bar \alpha_s\bigg(4\ln 2-14\zeta_3 \nu^2\bigg) \ ,
\end{align}
from which one has the standard
\begin{align}
\label{ABAB}
\alpha_{\rm BFKL}=4\ln2 \bar \alpha_s\ , \qquad {\cal D}_{\rm BFKL}=14\zeta_3\bar \alpha_s ,
\end{align}
with $\bar \alpha_s=\frac{\alpha_sN_c}{\pi}$.

Alternatively,  we may identify the tachyon propagator~(\ref{eq:stringpro}), with the BFKL   integral
\begin{align}
\int d\nu \nu \bigg(\frac{b_{10}b'_{10}}{b^2}\bigg)^{1+2i\nu}e^{\bar \alpha_s(4\ln 2-14\zeta_3 \nu^2)} \ ,
\end{align}
More specifically, the BFKL parameters are  now
\begin{align}
h+\bar h=1+2i\nu=\frac{d_\perp}{2}+\sqrt{-M_0^2-j-\frac{d_\perp^2}{4}} \ ,
\end{align}
with
\begin{align}
\alpha_{\rm BFKL}=4\ln2 \bar \alpha_s\ , \qquad {\cal D}_{\rm BFKL}= \frac{7}{2}\zeta_3\bar \alpha_s \ ,
\end{align}
replacing the identification (\ref{ABAB}).
This implies that the emergent AdS transverse dimensionality is
$D_\perp=d_\perp+1_z=3$,  leading to AdS$_{2+3=5}$. The ensuing BFKL evolution  is instead in an emergent AdS$_{5}$,
which is the identification made in~\cite{Stoffers:2012zw}. Namely, the relevant structure in the BFKL side is a product of two hyper-geometric functions,
 with the doubling of the pre-factors $\frac{zz'}{b^2}$.
\\
\\
{\bf Entanglement entropy and multiplicities at strong coupling:}
\\
The  string dual reduced entanglement density matrix in $1+1+D_\perp$ dimensions, can be derived explicitly for
long strings in the confining regime. Remarkably, its spectrum is dominated by the collective
eigenvalues~\cite{Liu:2018gae}

\begin{figure*}
\includegraphics[height=8cm,width=.49\linewidth]{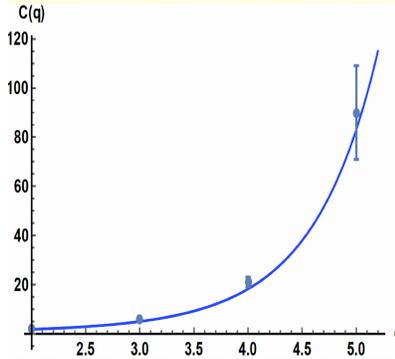}
\caption{Multiplicity moments extracted from $pp$ scattering at LHC at $\sqrt{s}=7$ TeV in the pseudo-rapidity
interval $\eta\leq 0.5$~\cite{CMS:2010qvf} as quoted in~\cite{Kharzeev:2017qzs}, versus the Polylog function following from the dual string analysis in (\ref{CQQ}).}
\label{fig_CQ}
\end{figure*}

\bea
\label{PDPERP}
p_n(D_\perp)=\frac{(n+D_\perp-1)!}{n!D_\perp!}e^{-\frac{D_\perp}6\,y}\bigg(1-e^{-\frac 16 y}\bigg)^n \ ,
\eea
which generalize (\ref{PNX}) to arbitrary $D_\perp$,  at strong coupling and largev $N_c$.
The remaining eigenvalues are small,
and  randomly (Poisson) distributed.  The normalized q-moments of (\ref{PDPERP}) are captured by a PolyLog
\bea
\label{CQQ}
C(q)=\frac {\langle n^q\rangle}{\langle n\rangle^q} =\frac 1{\langle n\rangle^q(\langle n\rangle -1)}\,{\rm PolyLog}\bigg(-q, 1-\frac 1{\langle n\rangle}\bigg) \ ,
\eea
with  the mean multiplicity
\bea
\langle n\rangle=D_\perp e^{\frac y6}\rightarrow D_\perp\bigg(\frac s{s_0}\bigg)^{\frac 16} \ .
\eea
The rightmost relation follows from the identification of the rapidity $y={\rm ln}{\frac s{s_0}}$, for hadron-hadron scattering at large $\sqrt{s}$.
In Fig.~\ref{fig_CQ}, we compare the  multiplicity moments extracted from $pp$ scattering at LHC at $\sqrt{s}=7$ TeV in the pseudo-rapidity
interval $\eta\leq 0.5$~\cite{CMS:2010qvf} as quoted  in~\cite{Kharzeev:2017qzs}, versus the moments $C(q)$ in (\ref{CQQ})  shown through the continuous q-Polylog.
The entanglement entropy is saturated by the collective eigenvalues of the string (\ref{PDPERP})
\bea
\label{STRINGENT}
-\sum_{n}p_n(D_\perp){\rm ln}\,p_n(D_\perp)\approx \frac {D_\perp}6\, y \rightarrow 2\alpha_{\mathbb P}\, y
\eea
with the  rightmost result following from (\ref{ALPHAP})~\cite{Stoffers:2012mn} (Note the difference of 1 in the definition of the Pomeron intercept in~\cite{Stoffers:2012mn}).  A similar result  was noted in~\cite{Zhang:2021hra},
 using a spin-chain analysis~\cite{Lipatov:1996ts}.
In diffractive $pp$ and DIS scattering, the entanglement entropy is captured by twice the Pomeron intercept, and  therefore {\it measurable}!
As we noted earlier, the additional partial tracing of (\ref{RHO1}),
may account for possible multiplicities with double rapidity gaps also in the diffractive regime,
and may be accessible in current $pp$   collisions at the LHC.

Finally, the correspondence with a string in AdS$_5$,  allows the identification of an Unruh-like temperature $T=(y/b)/2\pi$, on
the string world-sheet at large rapidity $y$,  and large $b$ with a wall to account for confinement~\cite{Basar:2012jb}. As a result, the effective thermal entropy associated to the
string using standard thermodynamics (derivative of the string classical free energy) at large rapidity~\cite{Stoffers:2012mn,Liu:2018gae}, is found to match the quantum
string entanglement entropy (\ref{STRINGENT}). This extends the concept of entanglement induced by small-$x$ radiation in quarkonium
at weak coupling $\alpha_s$ and large $N_c$, to strong $^\prime$t Hooft coupling $\lambda=g^2_sN_c$ and large $N_c$
in walled AdS$_5$ (a dual of confining QCD).

\section{Conclusions}~\label{sec_CONCLUSION}

Hadrons undergoing large boosts are surrounded by ever-growing wee parton clouds.
In perturbative QCD, this growth is dominated by soft gluon emission. This growth
is captured by the entanglement content of the hadronic wavefunction, with the
quarkonium wavefunction being the simplest illustrative example of this phenomenon.

We have explicitly constructed the quarkonium wavefunction on the light front at leading
order in $\alpha_s$,  by including both the real and virtual contributions. The
entanglement entropy from the real emission is found to be of order $\alpha_s y^2$,
in comparison to the virtual emission which is of order $\alpha_s y$. This
enhancement is  shown to propagate to higher orders, with a contribution of
 of $\alpha_s^ny^{n+1}$ to the entanglement entropy.

In the large $N_c$ limit and weak gauge coupling,  the leading density matrix of quarkonium,
obtained by tracing  over a single longitudinal cut $[x_{\rm min}, x_0]$, can be re-summed  in
a closed form.  The reduced density matrices with and without overall longitudinal momentum conservation,
are shown to obey non-local  BK-like integral equations for  QCD in any space-time dimension larger
than 2, as there is no radiative gluons in 2D. (They may generalize to strong gauge
coupling, using the arguments in~\cite{Kutak:2013hda}).

We solve these equations for QCD in 4D with only longitudinal evolution, and
non-conformal QCD in 3D including both transverse and longitudinal evolution. For the former,
the entanglement entropy is found to be at the Kolmogorov-Sinai bound of 1, when the in-out
longitudinal momenta are fixed. For the latter, the rate of change of the entanglement entropy
is found to vanish at large rapidities, since the soft gluon multiplication is limited by
a narrowing transverse space.

The rapidity evolution of the trace of the reduced density matrix without longitudinal
momentum conservation, obeys a diffusion like equation with BFKL kernels as noted
originally by Mueller~\cite{Mueller:1993rr,Mueller:1994gb}. This evolution maps onto an evolution
in an emergent AdS$_5$ space,  spanned by the $1+1$ longitudinal directions  plus additional
 transverse directions.

These observations extend to strong $^\prime$t Hooft coupling and large $N_c$, where the evolution of the partial  trace of the reduced
density matrix, is captured by the evolution of the tachyonic mode of a boosted string in AdS$_5$
space. The largest eigenvalues of the one-body reduced density matrix, gives a good account of the hadronic
multiplicities currently reported in $pp$ collisions at the largest $\sqrt{s}$ at the LHC. The eigenvalues
of the two-body reduced density matrix, may account for the multiplicities from $pp$ processes with a double
rapidity gap.

Finally, the boosted string  is characterized by an Unruh-like temperature
$T=(y/b)/2\pi$ on the world-sheet (with $b$ in arbitrary $D_\perp$ dimensions)~\cite{Basar:2012jb,Stoffers:2012mn}.
The string  effective thermal entropy, is the entanglement entropy  induced by small-$x$ gluons in quarkonium-quarkonium scattering,
extended to strong coupling. Feynman wee and perturbative partons at low-x~\cite{Feynman:1969wa}, are dual to Susskind-Thorn non-perturbative
string bits~\cite{Susskind:1993ki,Susskind:1993aa,Thorn:1994sw},  in a long string undergoing large boosts. For the latter,
the entanglement is captured geometrically by the hyperbolic string world-sheet~\cite{Liu:2018gae}.



\vskip 1cm
{\bf Acknowledgements}

This work was supported by the U.S. Department of Energy under Contract No.
DE-FG-88ER40388, and  by the Priority Research Area SciMat under the program
Excellence Initiative Research University at the Jagiellonian University in Krakow.

\appendix

\bibliography{ENT}

\end{document}